\input{aipcheck}
\documentclass[final,numberedheadings]{aipproc}
\usepackage{epsfig,latexsym}
\layoutstyle{6x9}
\begin{document}

\title{\boldmath
Inclusive electroproduction of light hadrons with large $p_T$ at
next-to-leading order
\unboldmath}


\author{Bernd A. Kniehl}{
  address={II. Institut f\"ur Theoretische Physik, Universit\"at Hamburg,
Luruper Chaussee 149, 22761 Hamburg, Germany}
}

\begin{abstract}
We review recent results on the inclusive electroproduction of light hadrons
at next-to-leading order in the parton model of quantum chromodynamics
implemented with fragmentation functions and present updated predictions for
HERA experiments based on the new AKK set.
\end{abstract}

\maketitle

\section{Introduction}

In the framework of the parton model of quantum chromodynamics (QCD), the
inclusive production of single hadrons is described by means of fragmentation
functions (FFs) $D_a^h(x,\mu)$.
At lowest order (LO), the value of $D_a^h(x,\mu)$ corresponds to the
probability for the parton $a$ produced at short distance $1/\mu$ to form a
jet that includes the hadron $h$ carrying the fraction $x$ of the longitudinal
momentum of $a$.
Analogously, incoming hadrons and resolved photons are represented by
(non-perturbative) parton density functions (PDFs) $F_{a/h}(x,\mu)$.
Unfortunately, it is not yet possible to calculate the FFs from first
principles, in particular for hadrons with masses smaller than or comparable
to the asymptotic scale parameter $\Lambda$.
However, given their $x$ dependence at some energy scale $\mu$, the evolution
with $\mu$ may be computed perturbatively in QCD using the timelike 
Dokshitzer-Gribov-Lipatov-Altarelli-Parisi (DGLAP) equations.
Moreover, the factorization theorem guarantees that the $D_a^h(x,\mu)$
functions are independent of the process in which they have been determined
and represent a universal property of $h$.
This entitles us to transfer information on how $a$ hadronizes to $h$ in a
well-defined quantitative way from $e^+e^-$ annihilation, where the
measurements are usually most precise, to other kinds of experiments, such as
photo-, lepto-, and hadroproduction.
Recently, FFs for light charged hadrons with complete quark flavour separation
were determined through a global fit to $e^+e^-$ data from LEP, PEP, and SLC
\cite{akk} thereby improving a previous analysis \cite{kkp}.

\begin{figure}[t]
\includegraphics[width=0.7\textwidth]{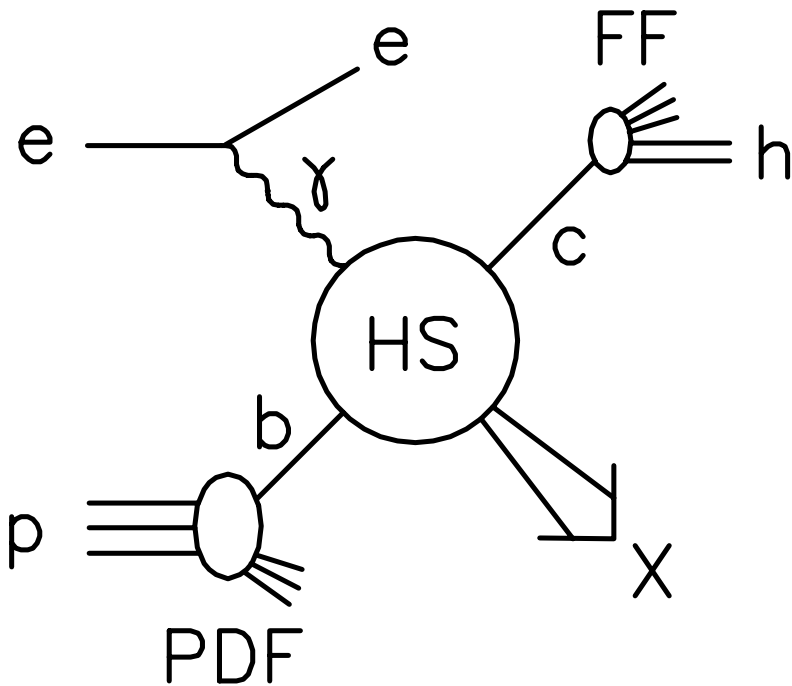} 
\caption{Parton-model representation of $ep\to eh+X$.}
\label{fig:pm}
\end{figure}
The QCD-improved parton model should be particularly well applicable to the
inclusive production of light hadrons carrying large transverse momenta
($p_T$) in deep-inelastic lepton-hadron scattering (DIS) with large photon
virtuality ($Q^2$) due to the presence of two hard mass scales, with
$Q^2,p_T^2\gg\Lambda^2$.
In Fig.~\ref{fig:pm}, this process is represented in the parton-model picture.
The hard-scattering (HS) cross sections, which include colored quarks and/or
gluons in the initial and final states, are computed in perturbative QCD.
They were evaluated at LO more than 25 years ago \cite{Mendez:zx}.
Recently, the next-to-leading-order (NLO) analysis was performed independently
by three groups \cite{Aurenche:2003by,kkm,Daleo:2004pn}.
A comparison between Refs.~\cite{kkm,Daleo:2004pn} using identical input
yielded agreement within the numerical accuracy.

The cross section of $e^+p\to e^+\pi^0+X$ in DIS was measured in various
distributions with high precision by the H1 Collaboration at HERA in the
forward region, close to the proton remnant \cite{Adloff:1999zx,Aktas:2004rb}.
This measurement reaches down to rather low values of Bjorken's variable
$x_B=Q^2/(2P\!\cdot\!q)$, where $P$ and $q$ are the proton and virtual-photon
four-momenta, respectively, and $Q^2=-q^2$, so that the validity of the
DGLAP evolution might be challenged by Balitsky-Fadin-Kuraev-Lipatov (BFKL)
dynamics.

In Ref.~\cite{kkm}, the H1 data \cite{Adloff:1999zx,Aktas:2004rb} were
compared with NLO predictions evaluated with the KKP FFs \cite{kkp}.
In Section~\ref{sec:two}, we present an update of this comparison based on the
new AKK FFs \cite{akk}.
Our conclusions are summarized in Section~\ref{sec:three}.

\section{Comparison with H1 data}
\label{sec:two}

We work in the modified minimal-subtraction ($\overline{\mathrm{MS}}$)
renormalization and factorization scheme with $n_f=5$ massless quark flavors
and identify the renormalization and factorization scales by choosing
$\mu^2=\xi[Q^2+(p_T^\ast)^2]/2$, where the asterisk labels quantities in the
$\gamma^\ast p$ center-of-mass (c.m.) frame and $\xi$ is varied between 1/2
and 2 about the default value 1 to estimate the theoretical uncertainty.
At NLO (LO), we employ set CTEQ6M (CTEQ6L1) of proton PDFs
\cite{Pumplin:2002vw}, the NLO (LO) set of AKK FFs \cite{akk}, and the
two-loop (one-loop) formula for the strong-coupling constant
$\alpha_s^{(n_f)}(\mu)$ with $\Lambda^{(5)}=226$~MeV (165~MeV)
\cite{Pumplin:2002vw}.

\begin{figure}[t]
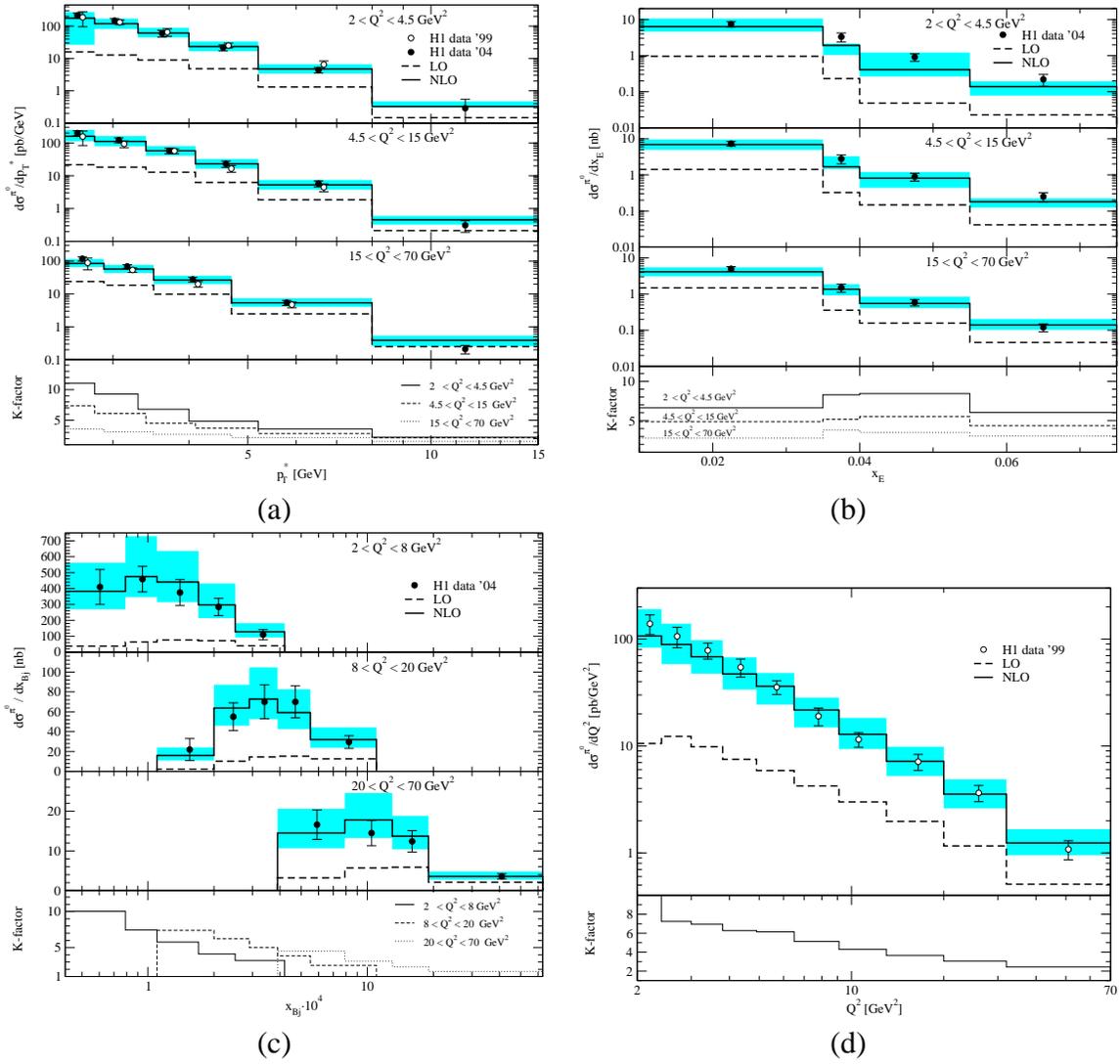

\begin{tabular}{cc}
\includegraphics[width=.5\textwidth]{dptAKK.eps} &
\includegraphics[width=.5\textwidth]{dEQqAKK.eps} \\
(a) & (b) \\
\includegraphics[width=.5\textwidth]{dxBjorken.eps} &
\includegraphics[width=.5\textwidth]{dQqAKK.eps} \\
(c) & (d)
\end{tabular}
\caption{H1 data on (a) $d\sigma/dp_T^*$, (b) $d\sigma/dx_E$, and (c) 
$d\sigma/dx_B$ for $2<Q^2<4.5$~GeV$^2$, $4.5<Q^2<15$~GeV$^2$, or
$15<Q^2<70$~GeV$^2$, and on (d) $d\sigma/dQ^2$ from Refs.~\cite{Adloff:1999zx}
(open circles) and \cite{Aktas:2004rb} (solid circles) are compared with our
default LO (dashed histograms) and NLO (solid histograms) predictions
including theoretical uncertainties (shaded bands).
The QCD-correction ($K$) factors are also shown.}
\label{fig:xs}
\end{figure}
The H1 data \cite{Adloff:1999zx,Aktas:2004rb} were taken in DIS of positrons
with energy $E_e=27.6$~GeV on protons with energy $E_p=820$~GeV in the
laboratory frame, yielding a c.m.\ energy of $\sqrt S=2\sqrt{E_eE_p}=301$~GeV.
The DIS phase space was restricted to $0.1<y<0.6$ and $2<Q^2<70$~GeV$^2$,
where $y=Q^2/(x_BS)$.
The $\pi^0$ mesons were detected within the acceptance cuts $p_T^*>2.5$~GeV,
$5^\circ<\theta<25^\circ$, and $x_E>0.01$, where $\theta$ is their angle with
respect to the proton flight direction and $E=x_E E_p$ is their energy in the
laboratory frame.
The comparisons with the our updated LO and NLO predictions are displayed in
Figs.~\ref{fig:xs}(a)--(d).

\section{Conclusions}
\label{sec:three}

We calculated the cross section of $ep\to e\pi^0+X$ in DIS for finite values
of $p_T^*$ at LO and NLO in the parton model of QCD \cite{kkm} using the new
AKK FFs \cite{akk} and compared it with a precise measurement by the H1
Collaboration at HERA \cite{Adloff:1999zx,Aktas:2004rb}. 

We found that our LO predictions always significantly fell short of the H1
data and often exhibited deviating shapes.
However, the situation dramatically improved as we proceeded to NLO, where our
default predictions, endowed with theoretical uncertainties estimated by
moderate unphysical-scale variations, led to a satisfactory description of the
H1 data in the preponderant part of the accessed phase space.
In other words, we encountered $K$ factors much in excess of unity, except
towards the regime of asymptotic freedom characterized by large values of
$p_T^*$ and/or $Q^2$.
This was unavoidably accompanied by considerable theoretical uncertainties.
Both features suggest that a reliable interpretation of the H1 data within the
QCD-improved parton model ultimately necessitates a full
next-to-next-to-leading-order analysis, which is presently out of reach,
however.
For the time being, we conclude that the successful comparison of the H1 data
with our NLO predictions provides a useful test of the universality and the
scaling violations of the FFs, which are guaranteed by the factorization
theorem and are ruled by the DGLAP evolution equations, respectively.

Significant deviations between the H1 data and our NLO predictions only
occurred in certain corners of phase space, namely in the photoproduction
limit $Q^2\to0$, where resolved virtual photons are expected to contribute,
and in the limit $\eta\to\infty$ of the pseudorapidity
$\eta=-\ln[\tan(\theta/2)]$, where fracture functions are supposed to enter
the stage.
Both refinements were not included in our analysis.
Interestingly, distinctive deviations could not be observed towards the lowest
$x_B$ values probed, which indicates that the realm of BFKL dynamics has not
actually been accessed yet.

\begin{theacknowledgments}
The author thanks G. Kramer and M. Maniatis for their collaboration.
This work was supported in part by BMBF Grant No.\ 05~HT1GUA/4.
\end{theacknowledgments}

\bibliographystyle{aipproc}   

\begin{thebibliography}{9}

\bibitem{akk} S.~Albino, B.~A.~Kniehl, and G.~Kramer,
Report No.\ DESY~05-022 and hep-ph/0502188, \emph{Nucl.\ Phys.\ B} (in press).

\bibitem{kkp} B.~A.~Kniehl, G.~Kramer, and B.~P\"otter,
\emph{Nucl.\ Phys.\ B} \textbf{582}, 514 (2000);
\emph{Phys.\ Rev.\ Lett.}\ \textbf{85}, 5288 (2000);
\emph{Nucl.\ Phys.\ B} \textbf{597}, 337 (2001).

\bibitem{Mendez:zx} A.~Mendez,
\emph{Nucl.\ Phys.\ B} \textbf{145}, 199 (1978).

\bibitem{Aurenche:2003by}
P.~Aurenche, R.~Basu, M.~Fontannaz, and R.~M.~Godbole,
\emph{Eur.\ Phys.\ J. C} \textbf{34}, 277 (2004).

\bibitem{kkm} B.~A.~Kniehl, G.~Kramer, and M.~Maniatis,
\emph{Nucl.\ Phys.\ B} \textbf{711}, 345 (2005);
\textbf{720}, 231(E) (2005).

\bibitem{Daleo:2004pn} A.~Daleo, D.~de Florian, and R.~Sassot,
\emph{Phys.\ Rev.\ D} \textbf{71}, 034013 (2005);
R.~Sassot, in these proceedings.

\bibitem{Adloff:1999zx} H1 Collaboration, C.~Adloff et al.,
\emph{Phys.\ Lett.\ B} \textbf{462}, 440 (1999).

\bibitem{Aktas:2004rb} H1 Collaboration, A.~Aktas et al.,
\emph{Eur.\ Phys.\ J.\ C} \textbf{36}, 441 (2004).

\bibitem{Pumplin:2002vw} J.~Pumplin, D.~R.~Stump, J.~Huston, H.-L.~Lai,
P.~Nadolsky, and W.-K.~Tung,
\emph{JHEP} \textbf{0207}, 012 (2002).

\end{thebibliography}

\end{document}